\documentclass[letterpaper,11pt]{article}

\usepackage{amsmath,amssymb}
\usepackage{array}

\begin{document}

\baselineskip=18pt

\newcommand{\eps}{\epsilon}
\newcommand{\pslash}{\!\not\! p}
\newcommand{\I}{\rm 1\kern-.24em l} 
\newcommand{\Tr}{\mathop{\rm Tr}}


\thispagestyle{empty}
\vspace{20pt}
\font\cmss=cmss10 \font\cmsss=cmss10 at 7pt

\begin{flushright}
CERN-PH-TH/2006-104\\
UAB-FT-603, SU-4252-829 \\
ROMA1-1433/2006
\end{flushright}

\hfill
\vspace{20pt}

\begin{center}
{\Large \textbf
{A custodial symmetry for $Z b\bar b$}}
\end{center}

\vspace{15pt}
\begin{center}
{\large 
Kaustubh Agashe$\, ^{a}$, Roberto Contino$\, ^{b, c}$, 
Leandro Da Rold$\, ^{d}$, Alex Pomarol$\, ^{d,e}$
} \vspace{20pt}

$^{a}$\textit{Department of Physics, Syracuse University, Syracuse, NY 13244, USA}
\\
$^{b}$\textit{Dipartimento di Fisica, Universit\`a di Roma ``La Sapienza'' and \\
INFN Sezione di Roma, I-00185 Roma, Italy}
\\
$^{c}$\textit{Department of Physics and Astronomy, Johns Hopkins University, \\
Baltimore, MD 21218, USA}
\\
$^{d}$\textit{IFAE, Universitat Aut{\`o}noma de Barcelona,
08193 Bellaterra, Barcelona, Spain}
\\
$^{e}$\textit{Theory Division, CERN, CH-1211 Geneva 23, Switzerland}
\end{center}

\vspace{20pt}
\begin{center}
\textbf{Abstract}
\end{center}
\vspace{5pt} {\small \noindent
We show that a subgroup of the custodial symmetry  O(3)
that protects $\Delta\rho$ from radiative
corrections  can also protect the $Zb\bar b$ coupling. 
This allows one to 
build models of electroweak symmetry breaking, such as
Higgsless, Little Higgs or 5D composite Higgs models,
that are safe from corrections to $Z\rightarrow b\bar b$.
We  show that  when this symmetry protects
$Zb\bar b$ it cannot simultaneously 
protect $Zt\bar t$ and $Wt\bar b$. 
Therefore one can expect to 
measure sizable deviations from the SM
predictions of these couplings at future collider experiments.
We also show under what circumstances 
$Zb_R\bar b_R$  can receive corrections in the right direction to 
explain the anomaly in the LEP/SLD forward-backward asymmetry $A^b_{FB}$.
}

\vfill\eject
\noindent


\section{Introduction}

One of the most elegant
solutions to the hierarchy problem is to consider that  
the Higgs boson, the scalar field responsible for electroweak symmetry breaking (EWSB),
is not  a fundamental particle.   
This approach is  clearly inspired by QCD, where  scalar  and pseudoscalar
states appear as composites of the strong dynamics.
In recent years there has been a revival of  interest in  this approach. 
The important new ingredient has been   
calculability, achieved by using either the idea of ``collective breaking''
\cite{LH} or extra dimensions.

As in the old technicolor~\cite{TC} or composite Higgs models~\cite{GK},
the main challenge of these new scenarios is to pass successfully   
all the  electroweak precision tests (EWPT). 
This is a non-trivial task, since in these theories
deviations from the Standard Model (SM) predictions usually  arise at the tree level
due to   mixing  effects between 
SM fields  and the heavy states of the new sector. 
One of the main difficulties   is to avoid large deviations in  the  $Zb_L\bar b_L$ coupling,
whose measured value is in
agreement with the SM prediction at the $0.25\%$ level.
This is difficult to overcome, since in these models the top, being heavy, couples  
strongly to the new sector. Since $b_L$ is in the same weak doublet  as $t_L$,
it usually suffers from large modifications to its couplings.

In this article we  will show that the custodial symmetry O(3), advocated long ago
to protect $\Delta\rho$~\cite{custodial}, can  
also  protect $Zb\bar b$.
In particular we will see that the $Zb_L\bar b_L$ coupling can be safe 
from corrections and at the same time the SU(2)$_L$-related couplings
$Zt_L\bar t_L$ and $Wt_L\bar b_L$ can receive sizable modifications. 
As an example, we will present the explicit calculations of these effects
in a 5D scenario of EWSB.
The custodial symmetry can also be used to protect the coupling  of the   $b_R$   
to the $Z$. However, the LEP and SLD experimental measurements 
of the forward-backward asymmetry $A_{FB}^b$
suggest that the coupling $Zb_R\bar b_R$ might deviate from its SM value.  
We will then study the possibility of having  large effects
in $Zb_R\bar b_R$ of the right magnitude and sign
as suggested by the experimental data.

Our analysis can 
be useful for any  scenario of EWSB that  contains a new sector
beyond the SM (BSM) invariant under the  global custodial symmetry.
This sector is defined to include the Higgs field as well.
Examples  are 
the  strongly  interacting sector of  technicolor  models, 
the extra fields added in Little Higgs theories
to avoid quadratic divergences,
or the bulk of a warped extra dimension present in    
some Higgsless~\cite{Csaki:2003zu} and 
composite Higgs~\cite{Contino:2003ve,Agashe:2004rs} models.

\section{The  coupling $Z \psi\bar   \psi$ and the custodial  symmetry}
\label{sec:custodial}

We will consider BSM  sectors with the following  
global symmetry breaking pattern~\cite{custodial}:
\begin{equation}
\text{O(4)}\rightarrow\text{O(3)}\, .
\label{custo}
\end{equation}
This breaking is equivalent to the more familiar custodial pattern
SU(2)$_L \otimes$SU(2)$_R\rightarrow$SU(2)$_V$ together with a  parity defined 
as the interchange $L\leftrightarrow R$ ($P_{LR}$).  
As we will see below, this discrete symmetry plays an important
role to protect the coupling of the $Z$ to fermions from non-zero  corrections.
The BSM sector also has to respect an SU(3)$_c\otimes$U(1)$_X$
symmetry corresponding to  the SM color group and an extra U(1) needed to fit
the hypercharges of the SM fields ($Y=T^3_R+X$). 
As usual \cite{custodial}, we will parametrize the symmetry breaking in Eq.~(\ref{custo})
by a  $2\times 2$  unitary matrix field $U$ transforming as a   $\bf{(2, 2)_0}$ under 
SU(2)$_L \otimes$SU(2)$_R\otimes $U(1)$_X$, whose VEV is 
given by $\langle U\rangle={\I}$.

Since the BSM sector is invariant under O(4),
we can  rotate to a basis in which  
each BSM field (or operator), ${\cal O}_{\rm BSM}$, 
has a definite  left and right isospin quantum number, 
$T_{L,R}$, and  its 3rd component,  $T^3_{L,R}$.
We will assume that every SM field $\Phi$ is coupled to a single BSM field (or operator): 
${\cal L}_{\rm int}=\Phi^\dagger  {\cal O}_{\rm BSM}+h.c.$.
This assumption is always fulfilled in the BSM models that we  are interested in.
It guarantees  that we can univocally 
assign to each SM field definite quantum numbers $T_{L,R}, T^3_{LR}$, corresponding
to those of  the operator ${\cal O}_{\rm BSM}$  to which it couples.
Notice that  this does not imply that the SM fields are in complete representations
of SU(2)$_L \otimes$ SU(2)$_R$, as it  is known not to be the case.

Let us  consider the implications of the custodial symmetry   
O(3)$=$SU(2)$_V\otimes P_{LR}$ on the 
coupling $Z\psi\bar\psi$, where $\psi$ denotes a generic SM fermion. 
At zero momentum,  this coupling is given by
\begin{equation}
\frac{g}{\cos\theta_W}\left[Q^3_L-Q\sin^2\theta_W\right]  Z^\mu \bar\psi\gamma_\mu\psi\, ,
\end{equation}
where $Q^3_L$ and $Q$ are respectively the 3rd-component SU(2)$_L$ charge and 
the electric charge of $\psi$.
Since the electric charge $Q$ is conserved, 
possible modifications to the coupling $Z\psi\bar\psi$ can only arise  from 
corrections to $Q^3_L$. Before EWSB we have $Q^3_L=T^3_L$, 
but this is not  guaranteed anymore after EWSB.
We will be interested only in non-universal corrections induced
by the BSM fields, and we will treat the SM $W_L^3$ field as
an external classical source which probes the left charge $Q^3_L$.
This is consistent since corrections induced through the renormalization
of the $Z$ kinetic term are universal.

We found two subgroups of the custodial symmetry SU(2)$_V\otimes P_{LR}$
 that can protect $Q^3_L$.
The first one is the subgroup 
U(1)$_L\otimes$U(1)$_R\otimes P_{LR}$  that it is  broken 
by $\langle U\rangle$ down to U(1)$_V\otimes P_{LR}$.
Although $P_{LR}$  is a symmetry of the BSM sector, it is not, in general,
respected by the coupling of $\psi$ to the BSM sector.
For $P_{LR}$ to be a symmetry also of ${\cal L}_{\rm int}=\bar\psi {\cal O}_\psi + h.c.$,
we must demand that  $\psi$ is an eigenstate of $P_{LR}$.
This implies
\begin{equation}
T_L=T_R\ ,\ \ \ \ \ \ T^3_R=T^3_L\, ,
\label{con1}
\end{equation}
for the field $\psi$. 
If  this is the case, the non-universal corrections to the charge $Q^3_L$ of $\psi$ are zero.
The proof goes as follows.
By U(1)$_V$ invariance, we have that   $Q^3_V=Q_L^3+Q_R^3$ is conserved, 
and therefore it cannot receive corrections:
\begin{equation}
\delta Q^3_V=\delta Q^3_L+\delta Q^3_R=0\, .
\label{conv}
\end{equation}  
On the other hand, by $P_{LR}$ invariance we have that the shift in $Q_L^3$ 
must be equal to the shift in~$Q_R^3$: 
\begin{equation}
\delta Q^3_L=\delta Q^3_R\,  .
\label{conplr}
\end{equation}  
Eq.~(\ref{conv}) and Eq.~(\ref{conplr}) imply that $\delta Q^3_L=0$. 
This proves that SM fermions that fulfill the condition (\ref{con1}) have 
their coupling to the $Z$  protected by the subgroup U(1)$_V\otimes P_{LR}$ of the 
custodial symmetry.

The second  example of a symmetry that can protect $Q^3_L$  
is that of the discrete transformation  
$|T_L,T_R;T^3_L,T^3_R\rangle\rightarrow |T_L,T_R;-T^3_L,-T^3_R\rangle$,
a subgroup of the custodial SU(2)$_V$.
We will denote this symmetry by   $P_{C}$.
Its action on  2-component spinors is given by $P_{C}=i\sigma_1$, while 
SO(3) vectors transform with $P_{C}=\text{Diag}(1,-1,-1)$.
According to our rule then, the SM $W^3_{L}$ can be assigned
an odd parity under $P_{C}$:  $W^3_{L}\rightarrow -W^3_{L}$.
For  $\psi$ to be  an eigenstate of this symmetry, it must have 
\begin{equation}
T^3_L=T^3_R=0\, .
\label{con2}
\end{equation} 
If this is the case, we have that $\delta Q^3_L=0$ at any order.
Indeed, if $\psi$ is an eigenstate of $P_C$, 
then $\bar\psi\gamma^\mu\psi$ is even under $P_C$   
and it cannot couple to $W_L^3$ that is odd.
Thus, the coupling of the $Z$ to SM fermions that fulfill Eq.~(\ref{con2}) 
is protected by the subgroup $P_C$ of the custodial symmetry.

It is important to notice that 
the symmetries discussed above can only protect the  coupling of the 
$Z$ to fermions at zero momentum. 
However, momentum dependent corrections to $Z\psi\bar \psi$
are parametrically suppressed in strongly coupled BSM sectors.
For example, in the case of $Zb_L\bar b_L$ a naive estimate gives 
$\delta g/g \sim (\lambda_t/g_{BSM})^2\; \xi_R^{-2}\, (q^2/\Lambda^2_{BSM})$,
where $\lambda_t \sim g_{BSM}\, \xi_L \xi_R$ is the top Yukawa coupling,
$\xi_{L}$ ($\xi_{R}$) is the degree of mixing between $t_{L}$ ($t_{R}$) and 
BSM states ($0 \leq \xi_{L,R} \leq 1$),
and $g_{BSM}$ is the coupling among the BSM particles.
Therefore, $\delta g/g$ can be sufficiently small for $g_{BSM}\gg\lambda_t$
(and $\xi_R$ not too small).

\section{Corrections to $Zb_L\bar b_L$
in custodial invariant models}

The symmetry argument given in the previous section shows how to
build Higgsless or composite Higgs models in which $Zb\bar b$ does not receive  
corrections from the BSM sector.
Let us start with the $Zb_L\bar b_L$ coupling.
In these models it has been commonly  assumed that
$b_L$ transforms as a  $\bf{(2,1)_{1/6}}$ representation of the
SU(2)$_L \otimes$SU(2)$_R\otimes$U(1)$_X$ group.
In that case, $b_L$ has 
the quantum numbers  $T_L=1/2$, $T_R=0$, $T_L^3=-1/2$ and $T^3_R=0$, which
fulfill neither the condition~(\ref{con1}) nor (\ref{con2}).
As a consequence, $Zb_L\bar b_L$ is not protected by the custodial symmetry.
Condition (\ref{con1}), however, suggests us 
a better assignment for the $b_L$ quantum numbers: 
\begin{equation}
T_L=1/2=T_R\, ,\quad  \text{and}\quad  T^3_L=-1/2=T^3_R\, .
\label{assignbl}
\end{equation}
This assignment guarantees 
that  $Z\bar b_Lb_L$ does not receive  corrections
from the BSM sector. Eq.~(\ref{assignbl})  implies  that
 $t_L$, being in the same SU(2)$_L$ doublet as $b_L$, has  to have the following 
assignments:  $T_L=T_R=1/2$ and $T_L^3=-T_R^3=1/2$.
Therefore, condition (\ref{con1}) is not satisfied  for $t_L$
and there will be corrections to the  $Zt_L \bar t_L$ coupling.
Similarly, the custodial symmetry  does not protect $W t_L\bar b_L$ (see below),  
and one can have large modifications in this coupling as well,
without affecting $Zb_L\bar b_L$.
At present, the  couplings of the top to the  gauge bosons   are not accurately measured.
Future accelerators, however,  will improve the measurements
of these couplings and  will be able to test this scenario.

\subsection{Operator analysis}

We give here an operator analysis for the couplings
of $q_L=(t_L,b_L)$ to the $Z$ and the $W$
based on the custodial symmetry. 
For  the assignment of Eq.~(\ref{assignbl}), we must  embed 
 $b_L$  in a $\bf{4}_{2/3}$ of O(4)$\otimes$U(1)$_X$, or, equivalently, 
\begin{equation}
 q_L\ \in\   {\bf{(2,2)_{2/3}}}\equiv Q_L
\label{bl}
\end{equation}
under SU(2)$_L \otimes$SU(2)$_R\otimes$ U(1)$_X$.
In addition to the SM doublet, this representation contains
an extra SU(2)$_L$ doublet $q^\prime_L$ that, not corresponding to  any SM field,
will play the role of a non-dynamical spectator.
We find two  single-trace dimension-4 operators that can contribute to the $Z$ couplings:
\begin{equation}
{\cal L}= c_1 \Tr\big[\bar Q_L\gamma^\mu Q_L \hat V_\mu]
 + c_2 \Tr\big[\bar Q_L\gamma^\mu V_\mu Q_L]\, ,
\label{opezbb}
\end{equation}
where $Q_L=\sigma^\alpha Q_L^\alpha$   is a $2\times 2$ matrix field,
\footnote{We use the basis $\sigma^\alpha=({\I},i\sigma_1,i\sigma_2,i\sigma_3)$ where
$\sigma_{a}$, $a=1,2,3$, are the Pauli matrices.}
$V_\mu=(i D_\mu U)U^\dagger$, $\hat V_\mu=(i D_\mu U)^\dagger U$, and
the covariant derivative is defined as 
$D_\mu U=\partial_\mu U+ig\sigma_aW^a_\mu U/2-ig^\prime B_\mu U\sigma_3/2$.
By imposing  $P_{LR}$,  under which 
 $U\rightarrow U^\dagger$,
$V_\mu\leftrightarrow \hat V_\mu$ and $Q_L \rightarrow \sigma^{\alpha\, \dagger} Q_L^\alpha$,
we obtain $c_1=c_2$.
There is also a double-trace operator that can  contribute to the  $Z$ coupling to $q_L$:
\begin{equation}
{\cal L}= c_3 \Tr\big[\bar Q_L \gamma^\mu i D_\mu U]\Tr[U^\dagger Q_L]+h.c.\, .
\label{opezbb2}
\end{equation}

To obtain the contributions  to $Zb_L \bar b_L$, $Zt_L\bar t_L$ and $Wt_L\bar b_L$ 
we plug
\begin{equation}
Q_L=\sigma_- b_L+\sigma_0 t_L+...\ ,\ \ \ U={\I}\ , \ \  \
  D_\mu U=\frac{ig\sigma_3 }{2 \cos\theta_W}
  Z_\mu +\frac{ig\sigma_+}{\sqrt{2}}W^+_\mu+...\, ,
 \end{equation}   
into  Eqs.~(\ref{opezbb})  and (\ref{opezbb2}),
where $\sigma_\pm=(\sigma_1\pm i\sigma_2)/2$ and  $\sigma_0=({\I}+\sigma_3)/2$. 
This  gives 
\begin{equation}
\frac{g}{\cos\theta_W}\left[\frac{c_2- c_1}{2}\ \bar b_L\gamma^\mu b_L -
 \frac{c_1+ c_2+ 2 c_3}{2}\ \bar t_L\gamma^\mu t_L \right]Z_\mu -
 \frac{g}{\sqrt{2}}\ (c_2+c_3)\, \bar t_L\gamma^\mu b_L W^+_\mu+h.c.\, .
\label{vff}
\end{equation}
As expected from the  symmetry argument, the contributions to $Zb_L \bar b_L$
vanish after imposing invariance under $P_{LR}$ ($c_1=c_2$), while
the contributions to the couplings of the top quark are different from zero.

The embedding  
of  $t_R$  in a  multiplet of SU(2)$_L \otimes$SU(2)$_R\otimes$ U(1)$_X$
is determined by the  top mass operator $\bar q_L U t_R$.
There are two possible invariant operators:
\begin{align}
& a)\quad \overline{\bf{(2, 2)}}_{\bf{2/3}}\bf{(2, 2)_{0}(1,1)_{2/3}}\, , &
& \text{or} &
& b)\quad \overline{\bf{(2, 2)}}_{\bf{2/3}}\bf{(2, 2)_{0}(1,3)_{2/3}}\, , 
\label{2embtR} 
\\
\intertext{implying respectively the two following embeddings for $t_R$:}
& a)\quad t_R \in {\bf (1,1)_{2/3}}\, , &
& \text{or} &
& b)\quad t_R \in {\bf (1,3)_{2/3} \oplus (3,1)_{2/3}}\, ,
\label{br}
\end{align}
which correspond respectively to a $\bf{1_{2/3}}$ and a $\bf{6_{2/3}}$ multiplet 
of O(4)$\otimes$U(1)$_X$.
In both cases $t_R$ has $T_L^3=T_R^3=0$, fulfilling the condition (\ref{con2}).
Therefore, its coupling to the $Z$ is protected by  the $P_C$ symmetry. 
\footnote{ For the  case (a) it is interesting to notice 
that   $t_R$ is  a singlet of the custodial symmetry and therefore
loop effects involving this field will not generate corrections to $\Delta\rho$.}
We can also perform an  operator  analysis for the $Z$ coupling to $t_R$. 
For the case (a), no invariant
operator can be written since $\Tr[V_\mu]=\Tr[\hat V_\mu]=0$.
For the case (b),  we have 
that $t_R$ corresponds to the $T^3_L=T^3_R=0$ component of
${\bf (1,3)_{2/3}}\equiv U_R$. 
There are two dimension-4 operators that can contribute to the $Z$ coupling to $t_R$:
\begin{equation}
{\cal L}= c_4 \Tr\big[\bar U_R\gamma^\mu U_R \hat V_\mu]+
 c_5 \Tr\big[\bar U_R\gamma^\mu \hat V_\mu U_R]\, .
\end{equation}
Using $U_R=\sigma_3 t_R+...$ we find that, as expected, the  contribution
to $Zt_R \bar  t_R$ vanishes.

In theories in which the Higgs arises as a pseudo-Goldstone boson (PGB) 
from the symmetry
breaking SO(5)$\rightarrow$ O(4),  
one has to embed the fermion multiplets into SO(5) representations.
We find two very simple options.
For the case~(a) we can use a $\bf{5_{2/3}}$ of SO(5)$\otimes$U(1)$_X$,  
that decomposes as
\begin{equation} \label{5rep}
\bf{5_{2/3}=(2, 2)_{2/3} \oplus (1,1)_{2/3}}
\end{equation}
under SU(2)$_L\otimes$SU(2)$_R\otimes$U(1)$_X$,
and contains the multiplets of Eqs.~(\ref{bl}) and (\ref{br}).
For the case~(b)  we can embed the top in a
$\bf{10_{2/3}}$:
\begin{equation} \label{10rep}
\bf{10_{2/3}=(2, 2)_{2/3} \oplus (1,3)_{2/3} \oplus (3,1)_{2/3}}\, .
\end{equation}
In the composite Higgs model of Ref.~\cite{Agashe:2004rs} the SM fermions were embedded  in 
spinorial representations of SO(5) ($\bf{4}$'s of SO(5)),
and the shift in the $Z b_L\bar b_L$ coupling implied severe bounds
on the masses of the new particles~\cite{Agashe:2005dk}.
By simply embedding the SM fields in either of the representations (\ref{5rep}), 
(\ref{10rep}), one can avoid large corrections to $Z b_L\bar b_L$ and
build successful composite Higgs models with a much lighter spectrum of new particles~\cite{prep}.

\subsection{Explicit  calculations in 5D models of EWSB}
\label{sec:5Dmodels}

In this section we focus on 5D composite Higgs models
realized in AdS$_5$ space-time~\cite{Contino:2003ve,Agashe:2004rs}, 
and compute the correction to $Z\psi\bar\psi$ induced by the first 
Kaluza-Klein (KK) mode.
In these theories the EWSB scale is given by $v=\epsilon f_\pi$,
where $f_\pi$ is the analog of the pion decay constant and 
$\epsilon$ is a model-dependent parameter bounded to be $0<\epsilon\leq 1$.
The experimental constraint from the Peskin-Takeuchi $S$ parameter 
generically requires $\epsilon\lesssim 0.5$.
Our result for $Z\psi\bar\psi$ will also apply 
to the class of Higgsless models in AdS$_5$~\cite{Csaki:2003zu}
after setting $\epsilon= 1$.

Let us denote with $c$ the fermion 5D bulk mass in units of the AdS curvature.
We will assume $-1/2<c<1/2$, since for $|c|>1/2$
the fermion zero modes are quite decoupled
from the 5D bulk and non-universal corrections to $Z\psi \bar\psi$
from the exchange of KK modes are exponentially suppressed
(this is the case for the first and second generation fermions).
There are two types of diagrams contributing to $Z\psi\bar\psi$, 
one involving the exchange of gauge KKs, the other involving fermionic KKs.
The contribution from the tower of SU(2)$_L\otimes$SU(2)$_R$ gauge KKs
is, at the tree level and at zero momentum:
\begin{equation}
\delta g\simeq \left(T^3_R-T_L^3\right)\frac{1-2 c}{2\sqrt{2}(3-2 c)}\,\epsilon^2\, ,
\label{gauge}
\end{equation}
where $\delta g(g/\cos\theta_W)  \bar\psi\gamma^\mu\psi Z_\mu$ gives the 
non-universal correction to the SM vertex.~\footnote{Eq.(\ref{gauge})
is valid for $-1/2\leq c < 1/2$. In the limit $c\to 1/2$ the same
formula applies with $(1-2c)\to 1/(\pi kR)$, where 
$\pi R$ is the proper length 
of the extra dimension and $k$ is the curvature of AdS$_5$.}
Effects from the fermion KKs are of the form
\begin{equation}
\delta g=\sum_{\rm KK}\sin^2\theta_{KK}\left(T^{3\, \rm KK}_{L}-T_L^3\right) \, ,
\label{fermion}
\end{equation}
where $\theta_{KK}$ is the mixing angle between the KK and $\psi$. 
This mixing occurs after EWSB and it is of order  
$\sin\theta_{KK} \sim \epsilon \sqrt{1/2-c}$.~\footnote{
This holds if all the KKs have similar masses of order $\Lambda_{BSM}$.
If the KK state mixing with $\psi$ has a smaller mass $m \ll \Lambda_{BSM}$,
then $\sin\theta_{KK}$ is larger by a factor ($\Lambda_{BSM}/m$).
}
Although the sum in Eq.~(\ref{fermion})  is over all the KK tower, 
a good approximation is obtained
by considering only the lowest mode.

In the case in which 
$q_L$ belongs to a $\bf (2, 2)_{2/3}$ of SU(2)$_L\otimes$SU(2)$_R\otimes$U(1)$_X$,
only fermionic KKs in the representations 
$\bf{(1,1)_{2/3}}$, $\bf{ (1,3)_{2/3} \oplus 
(3,1)_{2/3}}$ and $\bf{(3,3)_{2/3}}$
can mix with $b_L$ or $t_L$ at order~$\epsilon$. 
The coefficients of the operators in Eqs.~(\ref{opezbb}) and (\ref{opezbb2}) then read:
\begin{equation}
c_1=c_2\simeq \frac{1-2 c_q}{2\sqrt{2}(3-2 c_q)}\,\epsilon^2
+\frac{1}{2}\sin^2\theta_{KK}^{(1,1)}
+\frac{1}{2}\sin^2\theta_{KK}^{(3,1)} - \frac{3}{4} \sin^2 \theta^{(3,3)}_{KK}\
 ,\  \ \ c_3=0\, .
\label{cs}
\end{equation}
Here $\theta_{KK}^{(1,1)}$ is the mixing angle  between  $t_L$ and 
the KK in the $\bf{(1,1)_{2/3}}$ representation,
and $\theta_{KK}^{(3,1)}$ ($\theta^{(3,3)}_{KK}$) is the mixing angle between  $b_L$ and 
the KK in the $\bf{(3,1)_{2/3}}$  ($\bf{(3,3)_{2/3}}$)  representation.
In the case of a composite  Higgs model where $q_L$ is embedded in a ${\bf 5_{2/3}}$ 
of SO(5), the result is that of Eq.~(\ref{cs}) with only the gauge and $\bf (1,1)_{2/3}$
fermionic contributions turned on.
Eq.~(\ref{vff}) together with Eq.~(\ref{cs}) give us the tree-level
correction to the couplings of the $Z$ and the $W$ to the SM fermions $b_L$,$t_L$. 
Corrections of order $\sim \epsilon^2\sim 10-20\%$  are thus possible
if $q_L$ is strongly coupled to the 5D bulk dynamics
(\textit{i.e.} for $-1/2 < c \lesssim 0$), and they could be observed in future experiments 
that probe the couplings of the top quark.

\section{The coupling $Zb_R\bar b_R$}

The small ratio $m_b/m_t$ can be naturally explained
in the class of models under consideration by assuming
that the SM $b_R$ couples weakly to the BSM sector.
The shift in the coupling of $b_R$ to the $Z$ due to the BSM sector,
$\delta g_{Rb}$, will then be small.
This is the case, for example, when 
$q_L \in \bf{(2,2)_{2/3}}$ and
both $b_R$ and $t_R$ couple to the same BSM operator
transforming as a $\bf{ (1,3)_{2/3} \oplus (3,1)_{2/3} }$, case $(b)$ of Eq.~(\ref{2embtR}).

It is however interesting to consider the possibility that $b_R$ couples
more strongly to the BSM sector, 
since a positive shift $\delta g_{ Rb } \sim + 0.02$
would explain the $3 \sigma$ anomaly in the forward-backward
asymmetry $A^b_\text{FB}$ measured by the LEP and SLD 
experiments (see~\cite{LEPEWWG}).~\footnote{
A larger and negative shift, $\delta g_{Rb}  \sim - 0.17$ 
would also explain the data~\cite{Choudhury:2001hs},
but to obtain such a large shift
would require a very light spectrum of new particles.
We do not consider here this possibility.
}

If, for example, $b_R$ and $t_R$ couple to two different BSM operators, possibly with the same 
SU(2)$_L \otimes$SU(2)$_R \otimes$U(1)$_X$  quantum
numbers, then $m_b \ll m_t$ could follow from hierarchies in the couplings of the BSM sector.
In the case of the 5D models of section~\ref{sec:5Dmodels}
one can use Eqs.~(\ref{gauge}) and (\ref{fermion}) to calculate $\delta g_{ Rb }$.
For $b_R\in \bf{ (1,3)_{2/3}}$, only KK fermions in a $\bf{ (2,2)_{ 2/3 }}$ and 
$\bf{ (2,4)_{ 2/3 }}$ can mix with $b_R$ at order $\epsilon$. This gives,
for $|c_b| < 1/2$,
\begin{equation}
\delta g_{ R b } \simeq - \frac{ 1 - 2 c_b }{ 2 \sqrt{2}
( 3 - 2 c_b ) } \epsilon^2 - \frac{1}{2} \sin^2 \theta^{ (2,2) }_{ KK } + 
\sin^2 \theta^{ (2,4) }_{ KK }. 
\end{equation}
Here and in the following, $\theta_{KK}^{(r,s)}$ denotes the mixing angle between 
$b_R$ and the KK state with electric charge $-1/3$ 
in a  $\bf{ (r,s) }$ representation of SU(2)$_L \otimes$SU(2)$_R$ 
(if the representation $\bf{(r,s)}$ contains
more than one state with electric charge $-1/3$, 
then $\theta_{KK}^{(r,s)}$ will refer to the KK with $T_L^3 = -1/2$).
Thus, one can obtain a positive $\delta g_{ Rb }$ from the mixing of $b_R$ with the
KKs in the $\bf{ (2,4)_{ 2/3 } }$,
as needed to explain the $A_{ FB }^b$ anomaly.

A different possibility is that the SM $q_L$ itself couples to two different
BSM operators: the first responsible for generating the top mass, the second for
generating the bottom mass.~\footnote{An explicit realization of this
scenario in the context of a 5D composite Higgs model will be given in~\cite{prep}.} 
The coupling to this latter operator will in general
violate the custodial symmetry subgroup protecting $g_{Lb}$, 
but it is natural to assume that its coefficient is small, in order
to reproduce the small ratio $m_b/m_t$.
The resulting $\delta g_{ L b }$ will also be small, allowing at the same time 
a large coupling of $b_R$ to the BSM sector.
There are many choices for embedding $b_R$ in SU(2)$_L \otimes$SU(2)$_R\otimes$U(1)$_X$,
giving $\delta g_{ Rb }$ of either sign. The simplest choice is
\begin{equation}
b_R  \in \bf{ (1,1)_{-1/3} }\, ,
\end{equation}
which can be embedded in a $\bf{5}$ of SO(5).
In this case the BSM operator coupled to $q_L$
responsible for the bottom mass has to transform as a
$\bf{(2,2)_{-1/3}}$.
Since however $T^3_{L \, , R} = 0$ for $b_R$, 
the $P_C$ symmetry argument of section~\ref{sec:custodial} implies
$\delta g_{ Rb } = 0$ for both gauge and fermionic
contributions. 
Another possible choice is
\begin{equation}
b_R \in \bf{ (1,2)_{1/6} }\, ,
\end{equation}
which can be embedded into a $\bf{4}$ of SO(5). 
In this case the BSM operator coupled to $q_L$
can transform as either a
$\bf{(2,1)_{1/6}}$ or a $\bf{ (2,3)_{ 1/6} }$.
At order $\epsilon$, $b_R$ can mix with KKs in $\bf{ (2,1)_{1/6 } }$ and
$\bf{ (2,3)_{1/6} }$.
We find 
\begin{equation}
\delta g_{ R b } \simeq - 
\frac{ 1 - 2 c_b }{ 4 \sqrt{2}
( 3 - 2 c_b ) } \epsilon^2 - \frac{1}{2} \sin^2 \theta^{ (2,1) }_{ KK } 
+ \frac{1}{2} \sin^2 \theta^{ (2,3) }_{ KK } \, .
\end{equation}
Thus, one has $\delta g_{ R b} > 0$ from mixing with KKs in the $\bf{ (2,3)_{ 1/6 }}$,
as needed to explain the $A_{ FB }^b$ anomaly.
A few other examples with $1$ Higgs insertion
are indicated in Table~\ref{bRtable}.
\begin{table}[t]
\begin{center}
\setlength\extrarowheight{4pt}
\begin{tabular}{|| l | r @{} l | c ||}
\hline\hline 
\multicolumn{1}{||c|}{ $b_R$ }
& \multicolumn{2}{c|}{ $\delta g_{ Rb } |_{ gauge }/\Delta_g
  = -Q_A$  }
& $ \delta g_{ Rb } |_{ fermionic }$ 
\\[4pt] 
\hline
$\bf{ (1,3)_{ 2/3 } }$ 
& \hspace{2.2cm} $-$ & $1$ \hspace{2.2cm}
& $ - \frac{1}{2} \sin^2 \theta^{ (2,2) }_{ KK } + 
\sin^2 \theta^{ (2,4) }_{ KK }$
\\[2pt]
\hline 
$\bf{ (1,1)_{ -1/3 }}$
& & $0$
& $0$ 
\\[2pt]
\hline 
$\bf{ (1,3)_{ -1/3 } }$
& & $0$
& $0$ 
\\[2pt]
\hline
$\bf{ (1,2)_{1/6}}$
& $-1$ & $/2$ 
& $ - \frac{1}{2} \sin^2 \theta^{ (2,1) }_{ KK } 
+ \frac{1}{2} \sin^2 \theta^{ (2,3) }_{ KK } $
\\[2pt]
\hline
$\bf{ (1,2)_{-5/6 }}$
& $+1$ & $/2$ 
& $ \frac{1}{2} \sin^2 \theta^{ (2,1) }_{ KK } 
-  \frac{1}{4} \sin^2 \theta^{ (2,3) }_{ KK } $
\\[2pt]
\hline
$\bf{ (1,3)_{ -4/3 } }$
& $+$ & $1$ 
& $ \frac{1}{2} \sin^2 \theta^{ (2,2) }_{ KK } 
- \frac{1}{3} \sin^2 \theta^{ (2,4) }_{ KK } $
\\[2pt]
\hline\hline 
\end{tabular}
\end{center}
\caption{ \textit{
Several possible embeddings of $b_R$ in SU(2)$_L \otimes$SU(2)$_R\otimes$U(1)$_X$
and corresponding 
contributions to $\delta g_{Rb}$ from the first KK modes in 5D models
of EWSB: gauge contribution
(size and sign as given by the $b_R$ axial charge $Q_A=T_L^3-T_R^3$,
where we have defined $\Delta_g=\frac{1-2 c_b}{2\sqrt{2}(3-2 c_b)}\epsilon^2$),
and fermionic contribution.}}
\label{bRtable}
\end{table}

\section{Conclusions}

In models where the electroweak symmetry breaking is induced
by a new (strongly interacting) sector coupled to the SM fields, it is crucial 
for the new sector to  respect a custodial symmetry in order 
to prevent large corrections to $\Delta\rho$.
We have shown that the custodial symmetry O(3)
can also protect the $Zb_L\bar b_L$ coupling from
corrections.
This suggests that the custodial invariance might be a key ingredient
to build natural models of electroweak symmetry
breaking with a relatively light spectrum of new fermions, as required
by naturalness arguments.
A way to test this scenario is to look for modifications
in the couplings $Zt\bar t$, $Wt\bar b$, which cannot be protected
at the same time by the custodial symmetry and can receive
potentially large shifts.
Finally, we investigated the possibility of a modification of the
$Zb_R\bar b_R$ coupling, showing that a positive shift, as required
to explain the anomaly in the LEP/SLD forward-backward asymmetry $A_{FB}^b$,
is possible for certain choices of the $b_R$ custodial quantum numbers.

\section*{Acknowledgments}

A.P. thanks Antonio Delgado, Christophe Grojean and Riccardo Rattazzi for useful discussions.
The work of R.C. was partly supported by NSF grant P420-D36-2051.
The work of L.D. and A.P. was  partly supported   by the
FEDER  Research Project FPA2005-02211
and DURSI Research Project SGR2005-00916.
The work of L.D. was supported 
by the Spanish Education Office (MECD) under 
an FPU scholarship.
A.P. thanks the Galileo Galilei Institute for Theoretical Physics for 
hospitality and the INFN for partial support during the completion of this work.


\end{document}